\documentclass[12pt]{article}
\usepackage[]{epsfig}
\usepackage[centertags]{amsmath}
\usepackage{amsfonts}
\usepackage{amssymb}
\usepackage[english]{babel}
\usepackage{amsmath, amsthm, slashed,url}

\begin{document}

\title{\bf Singular soliton solution in the Chern-Simons-CP(1) model}
\author{
Lucas Sourrouille$^a$
\\
{\normalsize \it $^a$Departamento de F\'\i sica, FCEyN, Universidad
de Buenos Aires}\\ {\normalsize\it Pab.1, Ciudad Universitaria,
1428, Ciudad de Buenos Aires, Argentina}
\\
{\footnotesize  sourrou@df.uba.ar} } \maketitle

\abstract{We show that the Chern-Simons-CP(1) model can support a singular soliton solution in which the magnetic field is a Dirac delta. }



 \vspace{10mm}
\section{Introduction}
The two dimensional $CP(n)$ sigma model was introduced in the  late  seventies \cite{golo,golo1,golo2}, in the search of understanding the strong coupling effects in $QCD$. This model captures  several interesting properties, many of them present in four dimensional $QCD$\cite{witten,witten1,witten2,witten3}. Whereas in four dimensional $QCD$ is difficult to demonstrate the existence of these properties, in two dimensional $CP(n)$ sigma model it becomes comparatively simple. An important issue related to this type of models concern to the existence of soliton type solutions. For the simplest $CP(1)$ model topological solutions have been shown to exist\cite{polyakov}.  Nevertheless, the solutions are of  arbitrary size due to scale invariance. As argued originally by Dzyaloshinsky, Polyakov and Wiegmann\cite{polyakov1} a Chern-Simons term can naturally arise in this type of models and the presence of a dimensional parameter could  play some role stabilizing the soliton solutions. A first detailed consideration of this problem was done in Ref.\cite{voru} where a perturbative analysis around the scale invariant solutions (i.e no Chern Simons coupling $\kappa=0$) showed that  the solutions were pushed to infinite size. More recently, in Ref.\cite{my3}, a nonperturbative  analysis of the solutions was done, showing that the Chern-Simons-CP(1) system admit only trivial solutions in $\rm R^2$.

In this note we will show that the Chern-Simons-CP(1) model support a non-trivial solution if we define it in ${\rm R}^2\setminus D(0,\epsilon)$, where $D(0,\epsilon)$ is a disc centered at the origin and with an arbitrary radius $\epsilon$. We will show that our solution produce a magnetic field at $D(0,\epsilon)$ and that this magnetic field becomes a Dirac delta, in the limit $\epsilon \to 0$.

 \vspace{5mm}
\section{The model}
\label{sec}
We begin by considering a $(2+1)$-dimensional Chern-Simons model coupled to a complex two component field $n(x)$  described by the action

\begin{eqnarray}
S&=& S_{cs}+\int d^3 x |D_\mu n|^2
\label{S1}
\end{eqnarray}
Here  $D_{\mu}= \partial_{\mu} - iA_{\mu}$ $(\mu =0,1,2)$ is the covariant derivative and $S_{cs}$ is the Chern-Simons action given by

\begin{eqnarray}
 S_{cs}= \kappa\int d^3 x \epsilon^{\mu \nu \rho} A_\mu \partial_\nu A_\rho
\end{eqnarray}
where

\begin{eqnarray}
F_{\mu \nu}=\partial_{\mu}A_{\nu}-
\partial_{\nu}A_{\mu} \label{F}
\end{eqnarray}
The metric signature is $(1,-1,-1)$ and  the two component field $n(x)$ is subject to
the constraint $n^\dagger n = 1$.
The constraint can be introduced in the variational process with a Lagrange multiplier. Then we extremise the following action

\begin{eqnarray}
S&=& S_{cs}+\int d^3 x |D_\mu n|^2 + \lambda (n^\dagger n -1)
\label{}
\end{eqnarray}
The variation of this action yields the field equations

\begin{eqnarray}
D_\mu D^\mu n +\lambda n =0
\label{motion1}
\end{eqnarray}

\begin{equation}
\kappa\epsilon_{\mu \nu \rho} F^{ \nu \rho} = - J_\mu=i [n^\dagger D_\mu n - n(D_\mu n)^\dagger]
\label{motion}
\end{equation}
From the first of these equations we get $\lambda= -(n^\dagger D_\mu D^\mu n)$, so that

\begin{eqnarray}
D_\mu D^\mu n = -(n^\dagger D_\mu D^\mu n)n
\label{motion2}
\end{eqnarray}
The time component of Eq.(\ref{motion})
\begin{eqnarray}
2\kappa F_{12} =  -J_0 \label{gauss}
\end{eqnarray}
is  Gauss's law of Chern-Simons dynamics. Integrating it over the entire plane one obtains the important consequence that any object with charge $Q =\int d^2 x J_0$ also carries magnetic flux $\Phi = \int B d^2 x$ \cite{Echarge,E1,E2}:

\begin{eqnarray}
\Phi = -\frac{1}{2\kappa} Q,
\label{Q}
\end{eqnarray}
where in the expression of magnetic flux we renamed $F_{12}$ as $B$.

Defining the stress tensor as $T_{ \mu \nu}= (D_{\mu} n)^{\dagger} D_{\nu} n +(D_{\nu} n)^{\dagger} D_{\mu} n- g_{\mu \nu} \Big( (D_\eta n)^{\dagger} D^{\eta} n \Big)$, the energy functional for a static field configuration can be expressed as
\begin{eqnarray}
E=  \int d^2 x \Big(\kappa^2 B^2 + |D_i n|^2  \Big) \,,
\;\;\;\;\;\
i = 1,2  \label{statich}
\end{eqnarray}
The requirement of the finite energy solution implies the following boundary conditions
\begin{eqnarray}
\lim_{r \to \infty} n(x) = n^0 e^{i\alpha(\phi)}
\,,
\;\;\;\;\;\
\lim_{r \to \infty} A_i = \partial_i \alpha
\label{bound}
\end{eqnarray}
Here $n^0$ is a fixed complex vector with $(n^0)^\dagger n^0 = 1$ and $\alpha$ is common phase angle. This $\alpha$ depend on  $\phi$, the angle in coordinate space that parameterizes the boundary of the space. With these conditions the magnetic flux reads

\begin{eqnarray}
\Phi =& \int \,\,d^2 x B = \oint_{|x|=\infty} \,\, A_i dx^i  = 2\pi N,
\label{}
\end{eqnarray}
being $N$ is a topological invariant which takes only integer values. In order to evaluate the minimum of the energy, the expression (\ref{statich})  may be rewritten as

\begin{eqnarray}
E=\int d^2 x \,\, \Big( \kappa^2 B^2 +  |( D_1 \pm iD_2)n|^2  \mp B \Big) \pm \frac{1}{2} \oint J_j dx^j
\label{H2}
\end{eqnarray}
where we have used the identity
\begin{eqnarray}
|D_i n|^2 =   |( D_1 \pm iD_2)n|^2 \mp B \pm \frac{\epsilon^{ij}}{2} \partial_i J_j
\end{eqnarray}
In virtue of the boundary conditions (\ref{bound}), the line integral in (\ref{H2}) vanishes and we have

\begin{eqnarray}
E=\int d^2 x \,\, \Big( \kappa^2 B^2 +  |( D_1 \pm iD_2)n|^2  \Big) \mp \Phi
\label{H3}
\end{eqnarray}
This implies that the energy is bounded below by the magnetic flux (for positive flux we choose the lower signs and for negative flux we choose the upper signs):

\begin{eqnarray}
E \geq  |\Phi|
\label{H}
\end{eqnarray}
This bound is saturated by fields satisfying the first-order Bogomol'nyi self-duality equations\cite{bogo}.

\begin{eqnarray}
( D_1 \pm iD_2)n =0
\,,
\;\;\;\;\;\
B=0
\label{bogo}
\end{eqnarray}
Since static configurations that are stationary points of the energy are also stationary points of the action, the Euler-Lagrange equations of the theory will satisfied by the Bogomol'nyi self-duality equations (\ref{bogo}).

The solution of the field equations, in the static case, was recently studied in Ref.\cite{my3}. There, the authors consider the following ansatz with cylindrical symmetry

\begin{eqnarray}
n(\phi, r)=  \left( \begin{array}{c}
\cos(\frac{\theta(r)}{2})e^{i N \phi}\\
\sin(\frac{\theta(r)}{2} )\end{array} \right)
\,,
\;\;\;\;\;\
 A_\phi (r)= a(r)
\,,
\;\;\;\;\;\
A_r =0\,,
\label{ansatz}
\end{eqnarray}
and wrote the field equations in terms of this ansatz
\begin{eqnarray}
\partial_r^2 a(r)+\frac{\partial_r a(r)}{r}- \frac{a(r)}{r^2} - \frac{a(r)}{\kappa^2}=\cos^2(\frac{\theta(r)}{2})\frac{N}{r\kappa^2}
\label{m1}
\end{eqnarray}
\begin{eqnarray}
r\partial_r(r\partial_r \theta(r)) + N^2 \sin(\theta(r)) = -2Nra(r)\sin(\theta(r))\;,
\label{m2}
\end{eqnarray}
where the fields $\theta(r)$ and $a(r)$ are subject to the following boundary conditions

\begin{eqnarray}
\lim_{r \to 0} \theta(r) = \pi
\,,
\;\;\;\;\;\
\lim_{r \to 0} a(r) = 0
\label{b1}
\end{eqnarray}

\begin{eqnarray}
\lim_{r \to \infty} \theta(r) = 0
\,,
\;\;\;\;\;\
\lim_{r \to \infty}a(r) =-\frac{ N}{r}
\label{b2}
\end{eqnarray}
Then they analyzed, both numerically and analytically, the behavior of the solution in several discs of the plane ${\rm R^2}$, and show that the solution becomes trivial as the size of the disc becomes infinity. More specifically, they introduce a variable parameter $S$, being $S$ the radius of the diverse discs, so that the profile functions read as $\theta(r, S)$ and $a(r,S)$, and the boundary conditions (\ref{b1}) and (\ref{b2}) as

\begin{eqnarray}
\lim_{r \to 0} \theta(r,S) = \pi
\,,
\;\;\;\;\;\
\lim_{r \to 0} a(r,S) = 0
\label{b3}
\end{eqnarray}

\begin{eqnarray}
\lim_{r \to S} \theta(r,S) = 0
\,,
\;\;\;\;\;\
\lim_{r \to S}a(r,S) =-\frac{ N}{S}
\label{b4}
\end{eqnarray}
Then they were able to show that

\begin{eqnarray}
\lim_{S \to \infty}a(r,S) =0
\label{b5}
\end{eqnarray}
That is, the solution becomes trivial as the size of the disc is increased infinitely. This solution implies a zero magnetic field in the plane

\begin{eqnarray}
\lim_{S \to \infty}B(r,S) =0
\label{}
\end{eqnarray}
However, the magnetic flux on a disc does not depend on $S$

\begin{eqnarray}
\int B dx^2 &=& 2\pi \int_0^S dr \; r B= 2\pi \int_0^S dr \; r \frac{\partial_r (r a(r,S)}{r} =2\pi \left.r a(r,S)\right|_0^S
\nonumber \\
&=& 2\pi S\Big(-\frac{N}{S}\Big)=-2\pi N
\label{}
\end{eqnarray}
and therefore remains constant even though the size of the disc becomes infinite.

\section{The soliton solution}

Now, we are interested in exploring the the possibility of finding a solution of the field equations distinct from the trivial one. Before discussing the possible solutions of field equations, it is convenient to redefine the model (\ref{S1}) in a new region. In particular we are interested in excluding the point $(0,0)$ of our domain. In others words we define our theory in ${\rm R^2}\setminus D(0,\epsilon)$, where $D(0,\epsilon)$ is a close disc with center at the origin and radius $\epsilon$. More precisely, $D(0,\epsilon)$ is defined as

\begin{eqnarray}
D(0,\epsilon) = \lbrace x \in {\rm R^2} : d(0, x) \leq \epsilon \rbrace
\end{eqnarray}
With these considerations the ansatz (\ref{ansatz}) read as

\begin{eqnarray}
n(\phi, r)=  \left( \begin{array}{c}
\cos(\frac{\theta(r-\epsilon)}{2})e^{i N \phi}\\
\sin(\frac{\theta(r-\epsilon)}{2} )\end{array} \right)
\,,
\;\;\;\;\;\
 A_\phi (r-\epsilon)= a(r-\epsilon)
\,,
\;\;\;\;\;\
A_r =0\,,
\label{ansatz1}
\end{eqnarray}
and then the equations (\ref{m1}) and (\ref{m2}) becomes

\begin{eqnarray}
\partial_r^2 a(r-\epsilon)+\frac{\partial_r a(r-\epsilon)}{r-\epsilon}- \frac{a(r-\epsilon)}{(r-\epsilon)^2} - \frac{a(r-\epsilon)}{\kappa^2}=\cos^2(\frac{\theta(r-\epsilon)}{2})\frac{N}{(r-\epsilon)\kappa^2}
\label{m11}
\end{eqnarray}
\begin{eqnarray}
(r-\epsilon)\partial_r((r-\epsilon)\partial_r \theta(r-\epsilon)) + N^2 \sin(\theta(r-\epsilon)) = -2N(r-\epsilon)a(r-\epsilon)\sin(\theta(r-\epsilon))
\label{m22}
\end{eqnarray}
Since we are looking for a non trivial solution, we propose

\begin{eqnarray}
a(r-\epsilon)= -\frac{N}{r-\epsilon}
\label{gauge}
\end{eqnarray}
Note that this proposition implies a zero magnetic field in ${\rm R^2}\setminus D(0,\epsilon)$
\begin{eqnarray}
B=\frac{\partial_r((r-\epsilon)\,\,a(r-\epsilon))}{r-\epsilon}=0
\end{eqnarray}
Introducing the equation (\ref{gauge}) in (\ref{m11}) and (\ref{m22}) we have

\begin{eqnarray}
&&\frac{N}{r-\epsilon} = \cos^2(\frac{\theta(r-\epsilon)}{2})\frac{N}{r-\epsilon}
\nonumber \\[3mm]
&&(r-\epsilon)\partial_r((r-\epsilon)\partial_r \theta(r-\epsilon))=N^2 \sin(\theta(r-\epsilon))
\label{bogo4}
\end{eqnarray}
In order to solve this system we must establish the boundary conditions for $\theta$. As in (\ref{b2}) we propose at infinity
\begin{eqnarray}
\lim_{r \to \infty} \theta(r-\epsilon) = 0\;,
\label{b7}
\end{eqnarray}
However, in the other boundary we change the condition to

\begin{eqnarray}
\lim_{r \to \epsilon} \theta(r-\epsilon) = 2\pi
\label{b6}
\end{eqnarray}

It is important to note here, that since $r=\epsilon$ is a circle, the field $n(\phi, r)$ in the limit $r\to \epsilon$ is regular. Otherwise if $\epsilon=0$ the boundary condition (\ref{b6}) read as
\begin{eqnarray}
\lim_{r \to 0} \theta(r) = 2\pi
\label{}
\end{eqnarray}
and therefore
\begin{eqnarray}
\lim_{r \to 0} n(\phi, r) = \left( \begin{array}{c}
e^{i N \phi}\\
0\end{array} \right)
\label{}
\end{eqnarray}
So there are infinite possible limits in $r=0$ (one for each angle). We can check that the proposition (\ref{b6}) is consistent with the equations (\ref{bogo4}). For this, we expand to first order in power series of $\theta$ the function $\sin (\theta(r))$ and  $\cos(\frac{\theta(r)}{2})$ near $\theta = 2\pi$

\begin{eqnarray}
&&\sin (\theta(r-\epsilon)) \approx  \sin (2\pi) + \cos (2\pi) \frac{\Big(\theta- 2\pi\Big)}{2} = \frac{\Big(\theta- 2\pi\Big)}{2}
\nonumber \\[3mm]
&&\cos(\frac{\theta(r-\epsilon)}{2}) \approx \cos(\pi) - \sin(\pi)\frac{\Big(\theta- 2\pi\Big)}{4}  =-1
\label{}
\end{eqnarray}
so that the equation (\ref{bogo4}) is rewritten as

\begin{eqnarray}
&&\frac{N}{r-\epsilon} =\frac{N}{r-\epsilon}
\nonumber \\[3mm]
&&(r-\epsilon)\partial_r((r-\epsilon)\partial_r \theta(r-\epsilon))=N^2 \frac{\Big(\theta- 2\pi\Big)}{2}
\label{bogo5}
\end{eqnarray}
The first equation is trivial and can be eliminated. The solution of this system must be of the form

\begin{eqnarray}
\theta = 2\pi + c(r-\epsilon)^n\;,
\label{}
\end{eqnarray}
where is $c$ constant and $n$ a positive real number. Introducing this equation in (\ref{bogo5}) we have

\begin{eqnarray}
&&c\;n^2\; (r-\epsilon)^{n} = N^2 \frac{c\;(r-\epsilon)^n }{2}
\label{bogo6}
\end{eqnarray}
This equation demands that

\begin{eqnarray}
&&n = \frac{N}{\sqrt{2}}
\label{}
\end{eqnarray}
So, the solution behavior in the limit $r \to \epsilon$ should be of the form

\begin{eqnarray}
&&\theta = 2\pi + c(r-\epsilon)^{\frac{N}{\sqrt{2}}}\;,
\label{ap1}
\end{eqnarray}
In a similar form we can analyze the behavior of the solution for a large $r$. In this case the solution has the form

\begin{eqnarray}
\theta = b\;r^{-\frac{N}{\sqrt{2}}}\;,
\label{ap2}
\end{eqnarray}
where $b$ is a constant. Here, note that due to the symmetry of the equation (\ref{bogo4}) we only consider $N$ positive. It is important to remark that the equations (\ref{ap1}) and (\ref{ap2}) imply the existence of nontrivial soliton solution for the field $\theta$.

Consider, now, the line integral over a circle that enclose the disc $D(0,\epsilon)$

\begin{eqnarray}
\oint d {\bf l}\,\, {\bf A} =\int _0^{2\pi} d \theta \,\, r A_{\phi}= -2\pi N
\label{fluxo}
\end{eqnarray}

Since the magnetic field is zero in ${\rm R^2}\setminus D(0,\epsilon)$, this implies, if $N$ is distinct to zero, the existence of a magnetic field in $D(0,\epsilon)$. Also indicates that the magnetic flux is quantized at $D(0,\epsilon)$. For a given $N$ the magnetic flux is constant and do not depend on the size of $D(0,\epsilon)$. So, if $\epsilon$ becomes arbitrarily small, the magnetic field must become arbitrarily large in order to the magnetic flux remains constant. That is, the magnetic field becomes a Dirac delta as $\epsilon \to 0$. 
\begin{figure}
\centering
\includegraphics
[height=70mm]{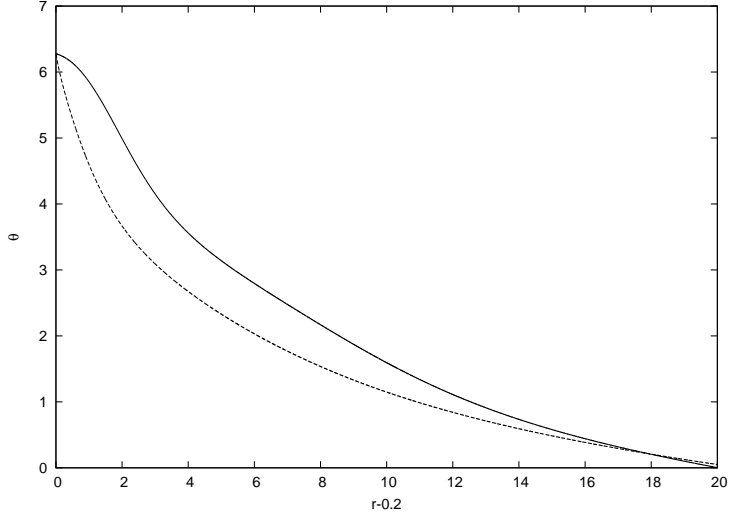}
\caption{{
Profile of $\theta$ for $\rm N=1$ (dashed line), $\rm N=2$ (solid line) and $\epsilon=0.2$
}}
\label{tt3}
\end{figure}

Note that if we propose, as in (\ref{b1}), the boundary condition

\begin{eqnarray}
\lim_{r \to \epsilon} \theta(r-\epsilon) = \pi
\label{con1}
\end{eqnarray}
instead of (\ref{b6}) and linearize the equation (\ref{bogo4}) in the limit $\theta \to \pi$ we obtain

\begin{eqnarray}
&&\frac{N}{r-\epsilon} = \frac{c^2 (r-\epsilon)^{2n}}{16} \frac{N}{r-\epsilon}
\nonumber \\[3mm]
&&c\;n^2\; (r-\epsilon)^{n} = -N^2 \frac{c\;(r-\epsilon)^n }{2}
\label{}
\end{eqnarray}
These equations have no solution, therefore (\ref{con1}) is not a good proposition for boundary condition.
It is also interesting to analyze the behavior of the energy at $r\to \epsilon$. In terms of the ansatz (\ref{ansatz1}) the energy (\ref{statich}) read
\begin{eqnarray}
E= 2\pi \int_{\epsilon}^{\infty} (r-\epsilon) dr \Big(\kappa^2 \left( \frac{a(r-\epsilon)}{r-\epsilon}+ \partial_r a(r-\epsilon) \right)^2 +\frac{1}{4}(\partial_r \theta(r-\epsilon))^2 \nonumber \\
+ \left(\frac{N^2}{(r-\epsilon)^2} + \frac{2Na(r-\epsilon)}{r-\epsilon}\right)\cos^2(\frac{\theta(r-\epsilon)}{2})+ a^2(r-\epsilon) \Big)
\;\;\;\;\;\
\label{}
\end{eqnarray}
Since, in our case we are considering $a(r)= -\frac{N}{r-\epsilon}$ this expression reduce to
\begin{eqnarray}
E= 2\pi \int_{\epsilon}^{\infty}(r-\epsilon) dr \Big(\frac{1}{4}(\partial_r \theta(r-\epsilon))^2
-\frac{N^2}{(r-\epsilon)^2} \cos^2(\frac{\theta(r-\epsilon)}{2})+\frac{N^2}{(r-\epsilon)^2}\Big)
\;\;\;\;\;\
\label{}
\end{eqnarray}
The expansion to a first order in $\theta$ of the $\cos^2(\frac{\theta(r)}{2})$ and the approximation $\theta = 2\pi + c\;(r-\epsilon)^n$, lead to following integrand near $\epsilon$
\begin{eqnarray}
2\pi (r-\epsilon){\cal E}= 2\pi  \frac{1}{4}(c\;n\;)^2 (r-\epsilon)^{2n-1}\;,
\label{}
\end{eqnarray}
where ${\cal E}$ is the energy density. This implies that $n \geq\frac{1}{2}$. Otherwise the energy diverges. Since in our solutions $n=\frac{N}{\sqrt{2}} \geq\frac{1}{2}$, this and the behavior of the fields at infinity ensure the finiteness of the energy solutions.

Finally, in Fig.\ref{tt3} we show numerical solutions for $N=1$ and $2$ in the case $\epsilon=0.2$.

\section{Conclusion}

In summary we have studied the classical solution of the Chern-Simons-CP(1) model defined on $\rm{R}^2\setminus D(0,\epsilon)$. For this, we propose as the solution of the gauge field
\begin{eqnarray}
a(r)= -\frac{N}{r-\epsilon}
\end{eqnarray}
and show that if $\theta$ is subject to the boundary condition

\begin{eqnarray}
\lim_{r \to \epsilon} \theta(r-\epsilon) = 2\pi\;,
\label{}
\end{eqnarray}
exists a non-trivial solution. Such solution has the particularity that produce a magnetic field at $D(0,\epsilon)$. In the limit case, as $\epsilon\to 0$ this magnetic field becomes a Dirac delta. While in Ref.\cite{my3} the authors considered the hole plane and showed that there are no solutions besides the trivial one, here we prove that excluding the origin of the plane it is possible to find a non trivial solution. It is, also, interesting to comment that similar solution for Yang-Mills-Higgs theory in (3+1) dimensions were found in references \cite{sin1}-\cite{sin2}-\cite{sin3}.

\vspace{1cm}
{\bf Acknowledgements}
I would like to thank Jer\'{o}nimo Peralta Ramos for reading the manuscript and many useful comments and suggestions. Also I thank Gaston Giribet for comments and D. Singleton for pointing out the existence of similar solution in Yang-Mills-Higgs theory.

\end{document}